\documentclass[aps,pre,preprint,superscriptaddress,showpacs]{revtex4}
\usepackage{graphicx,color}
\usepackage{amsmath,amssymb}


\begin{document}

\title{Stochastic switching in delay-coupled oscillators}
\author{Otti D'Huys}
\affiliation{Institute of Theoretical Physics, University of W\"urzburg, 97074 W\"urzburg, Germany}
\author{Thomas J\"ungling}
\affiliation{Instituto de Fisica Interdisciplinar y Sistemas Complejos, IFISC (UIB-CSIC),
Campus Universitat de les Illes Balears, E-07122 Palma de Mallorca, Spain}
\author{Wolfgang Kinzel}
\affiliation{Institute of Theoretical Physics, University of W\"urzburg, 97074 W\"urzburg, Germany}

\date{\today}
\begin{abstract}

A delay is known to induce multistability in periodic systems. Under influence of noise, coupled oscillators can switch between coexistent orbits with different frequencies and different oscillation patterns. For coupled phase oscillators we reduce the delay system to a non-delayed Langevin equation, which allows us to analytically compute the distribution of frequencies, and their corresponding residence times. The number of stable periodic orbits scales with the roundtrip delay time and coupling strength, but the noisy system visits only a fraction of the orbits, which scales with the square root of the delay time and is independent of the coupling strength. In contrast, the residence time in the different orbits is mainly determined by the coupling strength and the number of oscillators, and only weakly dependent on the coupling delay. Finally we investigate the effect of a detuning between the oscillators. We demonstrate the generality of our results with delay-coupled FitzHugh-Nagumo oscillators.

\end{abstract}

\maketitle


\section{Introduction}
In recent years dynamical systems with delays have evolved as a major topic in nonlinear sciences \cite{bookThomas,E}. Time delays arise naturally, and might play a role in many areas of physics, biology and technology, such as nonlinear optics \cite{RMPIngo,Nix12}, gene regulatory circuits \cite{Chen2002}, population dynamics \cite{NEL02,Mie12}, traffic flows \cite{Orosz04,Saf02}, neuroscience \cite{C}, and social or communication networks \cite{WAN06,Lu11}. 

A well established effect of a delay in the dynamics is the possibility to induce multistability \cite{Mas02.1,Kim97}. In oscillatory systems a delay gives rise to coexistent periodic orbits with different frequencies \cite{sch89.1,Yanchuk09,Sie13} and possibly different oscillation patterns \cite{ikke,choe10,Per10,Wil13,Vuellings14}. Such coexistent patterns could be related to memory storage and temporal pattern recognition, especially in neural networks \cite{Fos96,Fos97,RMPRab}. However, noise, which is unavoidable in real networks, can place important limitations to the capacity of a memory  element, as it can induce mode hoppings between coexistent attractors. 

We study the statistical properties of such mode hoppings in small networks of oscillators. We consider a single oscillator with delayed feedback, two delay-coupled oscillators and a unidirectional ring, and we briefly discuss globally coupled elements. The number of possible frequencies scales with the roundtrip delay time, but the noisy system visits only a fraction of these frequencies, which scales with the square root of the delay time. While without noise the range of frequencies also scales with the coupling strength, we find that in the stochastic system it does not depend on the coupling strength. In contrast, the robustness of the orbits to noise, measured by the average residence time, is mainly determined by the coupling strength, while the delay has a minor effect. Complementary to local stability analysis, the study of coupled systems subject to noise also provides information about the robustness of certain oscillation patterns. We find that depending on network topology, an oscillation pattern might dominate: in unidirectional rings the oscillators spend equally much time in all the possible phase configurations, a globally coupled network shows a clear preference to in-phase synchrony. 

This paper is organized as follows. In section II we discuss stochastic switching for a single phase oscillator. We compare our numerical results to those obtained by a potential model \cite{Mork90b}, and discuss the model in the limit of strong coupling and large delay. We discuss stochastic switching of two coupled phase oscillators in section III, and extend the potential model. In section IV we extend our results to a unidirectional ring of delay-coupled oscillators. Finally, we demonstrate the generality of our results with delay-coupled FitzHugh-Nagumo oscillators in section V. We discuss our results in section VI.


\section{Stochastic switching in a single phase oscillator with feedback} 

The most basic delay network is a single oscillator with delayed feedback. We consider a Kuramoto oscillator, which describes the oscillating dynamics by a single phase variable. It is a universal model, as many oscillators can be reduced to phase oscillators in the weak coupling regime \cite{kur97.1,dai97.1,AcebronKuramoto}. Thanks to its simplicity, the Kuramoto model allows for analytical insights while still capturing many essential features of synchronization. A Kuramoto oscillator with delayed feedback and noise is modelled by
\begin{equation}
\dot{\phi}(t)=\omega_0+\kappa\sin(\phi(t-\tau)-\phi(t)+\theta)+\xi(t)\label{eq:model}\,.
\end{equation}
The oscillator has a natural frequency $\omega_0$, the other parameters are the coupling delay $\tau$, the coupling strength $\kappa>0$ and the coupling phase $\theta$. The system is subject to additive Gaussian white noise $\xi(t)$, with $\langle\xi(t)\xi(t_0)\rangle=2D\delta(t-t_0)$. As the dynamics is invariant under a transformation $\phi(t)\rightarrow \phi(t)+\tilde{\omega}t,\omega_0\rightarrow \omega_0+\tilde{\omega},\theta\rightarrow\theta-\tilde{\omega}\tau$, we can omit the coupling phase $\theta$ without loss of generality. 

\begin{figure}[!ht]
\includegraphics[width=0.5\columnwidth]{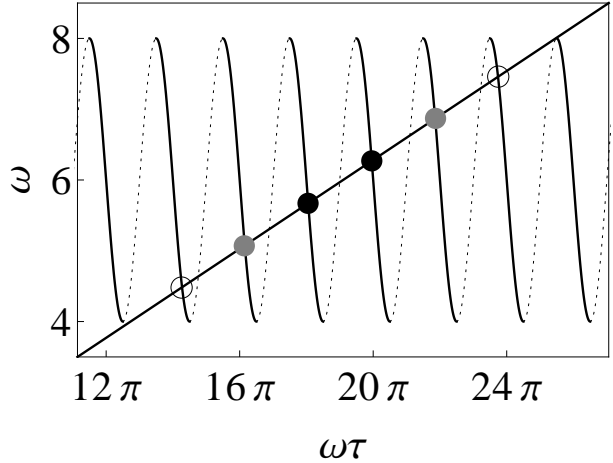}
\caption{Graphical determination of the different coexisting frequencies of a single oscillator with delayed feedback (Eq. \eqref{eq:omega}). The intersections with the thick decreasing slopes of the sine function correspond to stable orbits, and are marked with a circle. The coloring of the circles relates to the probability distribution $p(\omega(t))$ of the corresponding stochastic oscillator (shown in Fig. \ref{fig:var}): the probability that the oscillator has a frequency $\omega(t)\approx \omega_k$ is large for the most central frequencies $\omega_k\approx\omega_0$, marked with a black circle, while the probability to find the system's frequency $\omega(t)$ close to the outer frequencies $\omega_k\approx\omega_0\pm\kappa$, marked with an empty circle, is negligible. Parameters are $\omega_0=6$, $\kappa=2$, $\tau=10$ and $D=0.5$}
\label{fig:kura1}
\end{figure}

We first briefly discuss the deterministic dynamics of this system \cite{ear03.1,AmpPhase}. Without noise, the oscillator resides in one of the frequencies $\dot{\phi}=\omega_k$ given by
\begin{equation}
\omega_k=\omega_0-\kappa\sin(\omega_k\tau)\label{eq:omega}\,.
\end{equation}
A graphical determination of the frequencies $\omega_k$ is shown in Fig. \ref{fig:kura1}. The orbits for which $\kappa\tau\cos(\omega_k\tau)>1$ holds, are stable. For large coupling or long feedback delay $\kappa\tau\gg1$, the stable frequencies close to $\omega_0$ are approximated as $\omega_k\tau\approx 2k\pi$, whereas the spacing is given by $\omega_{k+1}-\omega_k\approx2\pi/\tau$. As all solutions of Eq.~\eqref{eq:omega} are limited by $\omega_0-\kappa\le\omega_k\le\omega_0+\kappa$, the number of coexistent stable orbits is estimated as $\kappa\tau/\pi$.

\begin{figure}[!ht]
\includegraphics[width=\columnwidth]{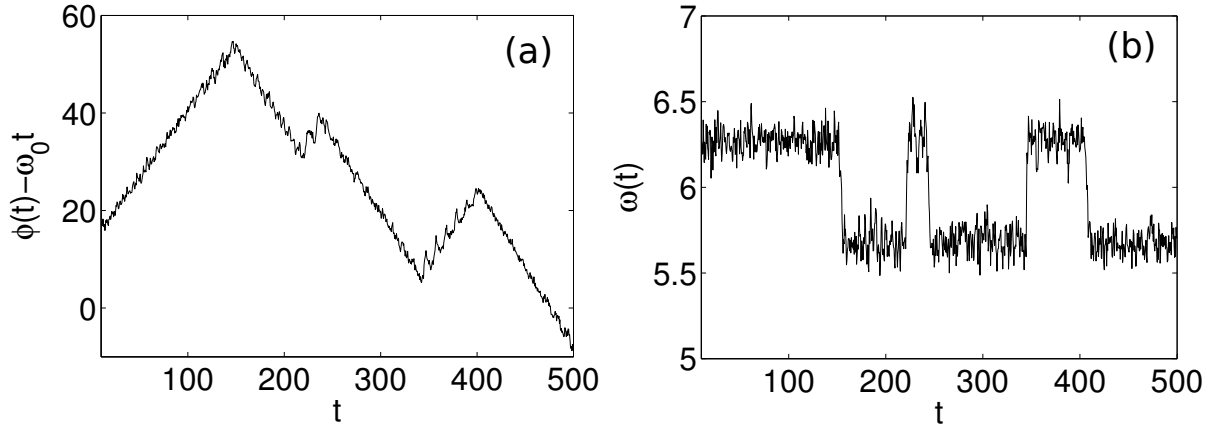}
\caption{(a) Phase evolution $\phi(t)-\omega_0 t$ of a Kuramoto oscillator with delayed feedback and noise. We subtracted the natural frequency $\omega_0 t$ for better visibility of the mode hoppings. (b) The frequency measure $\omega(t)=(\phi(t)-\phi(t-\tau))/\tau$ is a good indicator for the mode hoppings. Parameters are $\omega_0=6$, $\kappa=2$, $\tau=10$ and $D=0.5$}
\label{fig:kura1tt}
\end{figure}

If we add noise to the system, the oscillator switches between these coexistent orbits. We simulated a Kuramoto oscillator with delayed feedback, using a Heun algorithm adapted to delayed interactions, with a timestep of $h=0.01$. For our choice of parameters ($\kappa=2$, $\omega_0=6$, $\tau=10$), without noise, the oscillator has six stable periodic orbits, with respective frequencies $\omega_1\approx 4.48$, $\omega_2\approx 5.07$, $\omega_3\approx 5.67$, $\omega_4\approx 6.27$, $\omega_5\approx 6.87$ and $\omega_6\approx 7.46$, shown in Fig. \ref{fig:kura1}. A typical timetrace of the phase evolution, with multiple mode hoppings between $\omega_3$ and $\omega_4$, is shown in Fig. \ref{fig:kura1tt}(a). As an indicator for mode hoppings we use the frequency measure $\omega(t)=(\phi(t)-\phi(t-\tau))/\tau$, which is the driving term of the dynamics, and corresponds to the average of the instantaneous frequency $\dot{\phi}(t)$ over the past delay interval. Moreover, this definition of $\omega(t)$ respects the origin of the frequency locking, which lies in the auto-phase locking of the instantaneous phase $\phi(t)$ onto the delayed phase $\phi(t-\tau)$. The time evolution of $\omega(t)$ is shown in Fig.~\ref{fig:kura1tt}(b), exhibiting clear jumps between the deterministic frequencies $\omega_k$.

\begin{figure}[!hb]
\includegraphics[width=\columnwidth]{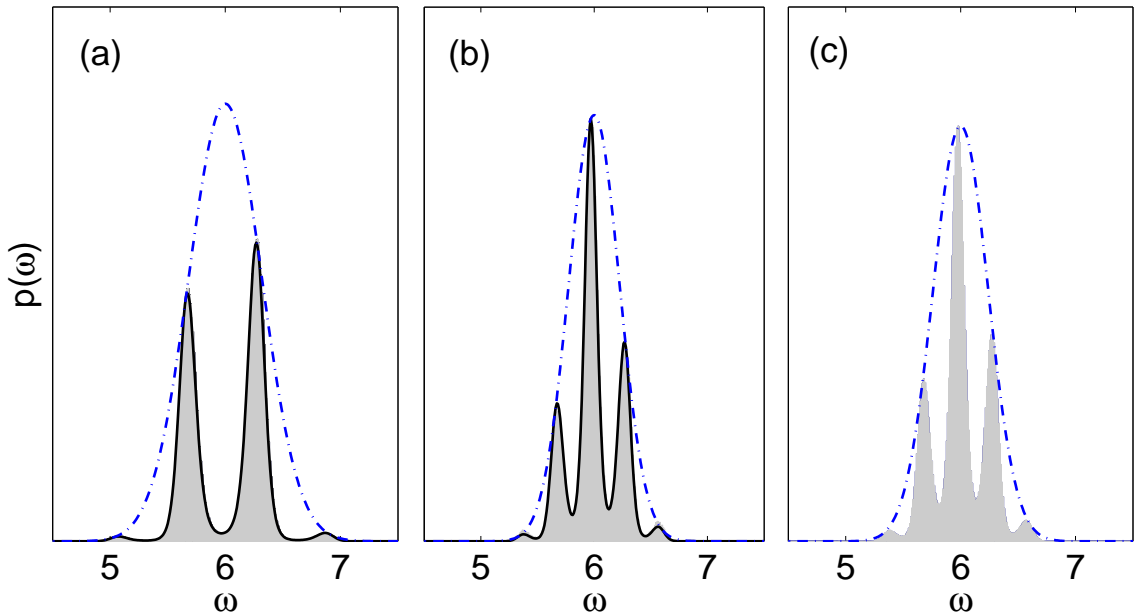}
\caption{(Color online) The frequency distributions (grey) for (a) an oscillator with feedback, (b) two coupled identical identical oscillators and (c) two detuned oscillators. The analytical approximations (Eq. \eqref{eq:frequencydistribution} and Eq. \eqref{eq:freqdistr2}) are plotted in black, the (blue) dashdotted lines show the respective Gaussian envelopes. The parameters are $\kappa=2$, $\tau=10$, $\omega_0=6$, $D=0.5$ , and (c) $\Delta=0.8$.}
\label{fig:var}
\end{figure}

The distribution of frequencies $p(\omega(t))$, with $\omega(t)$ defined as above, is shown in Fig. \ref{fig:var}(a). One can clearly distinghuish multiple maxima, corresponding to the deterministic frequencies $\omega_2$, $\omega_3$, $\omega_4$ and $\omega_5$ . The frequencies closest to the eigenfrequency $\omega_0$ of the oscillator, i.e. $\omega_3$ and $\omega_4$, are most often visited, while the oscillator spends a negligible amount of time in the orbits with frequencies $\omega_{1,6}\approx \omega_0\mp\kappa$.

\begin{figure}[!th]
\includegraphics[width=\columnwidth]{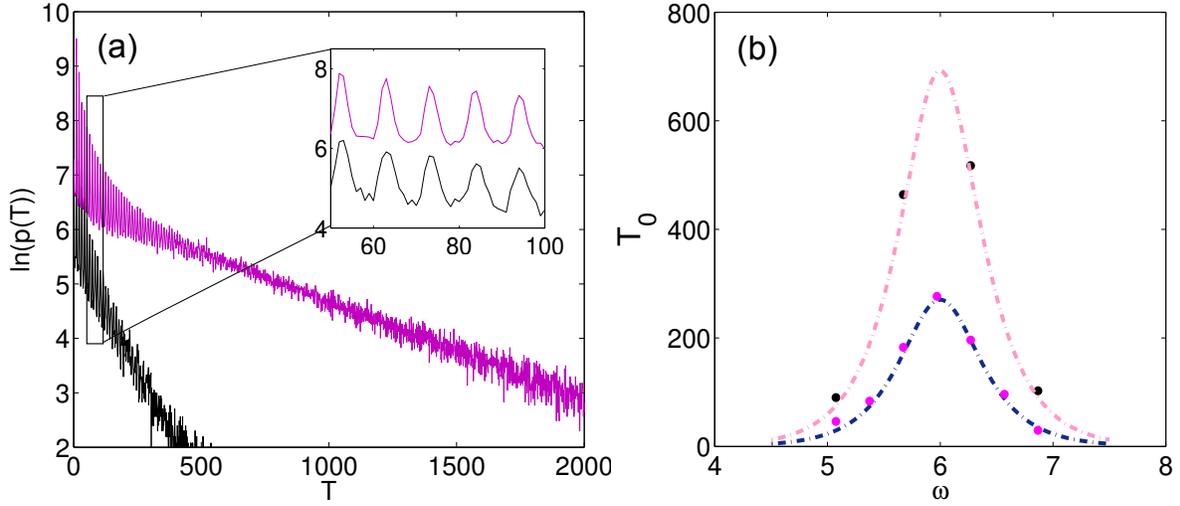}
\caption{(Color online)(a) Logarithm of the residence time distribution $\ln(p(T))$, for a Kuramoto oscillator with delayed feedback, for the orbits with frequencies $\omega_2$ and $\omega_3$. (b) Mean residence time of the orbits $\omega_{2,3,4,5}$ versus their frequency for a single oscillator (upper black dots) together the theoretical approximation (Eq. \eqref{eq:rtd1}) (upper dashed pink curve). The lower pink dots and the lower blue dashed curve represent the mean average residence times of the orbits and their theoretical approximation (Eq. \eqref{eq:rtd2}) respectively for two identical coupled systems. The parameters are $\kappa=3$, $\tau=10$, $\omega_0=6$ and $D=0.5$.}
\label{fig:rtd}
\end{figure}

To calculate the residence times of the orbits, we apply the following procedure: at the starting point $t_0$ the oscillator is considered to reside in the orbit with a frequency $\omega_k$ for which the distance $|\omega(t_0)-\omega_k|$ is minimal, and it stays there as long as $|\omega(t)-\omega_k|<\epsilon$. After a transition, we determine the new locking frequency again as the frequency at minimal distance. We chose $\epsilon=2/3(\omega_k-\omega_{k-1})$; for weak noise $\omega(t)$ does not show large fluctuations around the locking frequency $\omega_k$ and the choice of $\epsilon$ does not largely affect the residence times of the orbits. In our simulations we obtained around $10^6$ transitions. The residence time distributions of two of the orbits ($\omega_2$ and $\omega_3$) are shown in Fig. \ref{fig:rtd}(a). The distribution is exponential. Upon the exponential decay there are signatures of the delay time; these are shown in the inset. The peaks can be understood as delay echoes which result from a known stochastic resonance effect in delay systems~\cite{Ohi99,Mas02.1,Mas02.2}: A mode hopping causes a perturbation, which increases the probability for a mode hopping at multiples of the feedback delay. Moreover, the average residences times, shown in Fig. \ref{fig:rtd}(b), are largest for orbits $\omega_3$ and $\omega_4$ with a frequency close to the natural frequency $\omega_0$.

In order to interpret the mode hopping dynamics, we approximate the delay system by an undelayed system. It is then possible to define a Langevin equation and to compute the frequency distributions and average residence times of the different periodic orbits. Such an approach is possible thanks to the simplicity of the Kuramoto oscillator, as the dynamics of the oscillator is only characterized by a frequency. A similar method has been suggested in the context of mode hopping between external cavity modes in a single laser with delayed feedback \cite{Mork90b,Lenstra91}. Using this approximation, we show analytically how the frequency distribution and average residence times scale with the feedback strength, delay, and frequency of the orbit. Thereby we focus on the regime $\kappa\tau\gg 1$, in which a multitude of orbits coexists. 

\begin{figure}[!h]
\centering
\includegraphics[width=0.6\columnwidth]{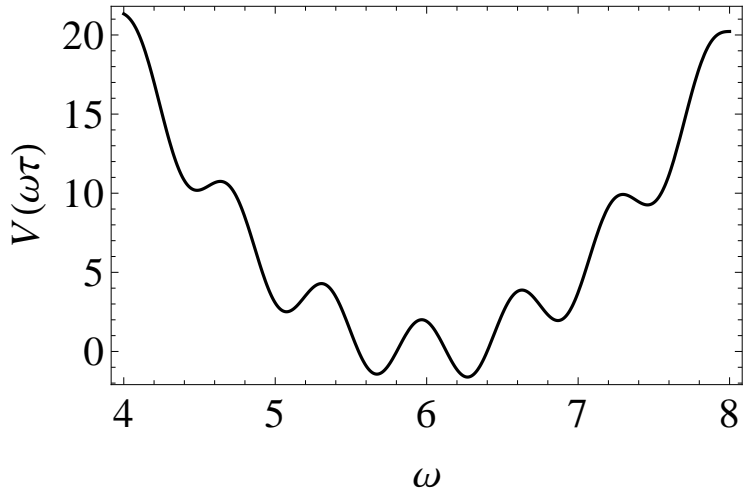}
\caption{Potential for a Kuramoto oscillator with feedback (Eq. \eqref{eq:potential}). Parameters are $\omega_0=6$, $\kappa=2$, $\tau=10$.}
\label{fig:kura1pot}
\end{figure}

In order to simplify the system, we first rewrite the dynamics in terms of the delay phase difference $x(t)=\phi(t)-\phi(t-\tau)$.
\begin{equation}
\dot{x}(t)=\omega_0-\kappa\sin x(t)-\dot{\phi}(t-\tau)+\xi(t)\,.
\end{equation}
The main step is the following: We approximate the instantaneous frequency $\dot{\phi}(t-\tau)$ by the frequency averaged over the future delay interval plus its noise source
\begin{equation}
\dot{\phi}(t-\tau)\approx\frac{1}{\tau}\int_{t-\tau}^{t}\dot{\phi}(t')dt'+\xi(t-\tau)=\frac{x(t)}{\tau} + \xi(t-\tau)\,.
\end{equation}
Such assumption is justified for weak noise, when the oscillator resides in one of the periodic orbits during a delay interval. But also in case of a random walk ($\kappa=0$) it leads to the correct stationary distribution. In this way we obtain a closed equation without delay for the phase difference $x(t)$, that can be written in terms of a potential \cite{Mork90b},
\begin{eqnarray}
\dot{x}(t) & = & -\frac{dV(x)}{dx} + \tilde{\xi}(t)\mbox{ with }\nonumber\\
V(x) & = & \frac{1}{2\tau}(x-x_0)^2 - \kappa\cos x\label{eq:potential}\,,
\end{eqnarray}
\noindent with $x_0=\omega_0\tau$ and $\tilde{\xi}(t)=\xi(t)-\xi(t-\tau)$. As the noise sources $\xi(t)$ and $\xi(t-\tau)$ are uncorrelated, the simplified oscillator is effectively subject to a magnified noise strength of $\langle\tilde{\xi}^2(t)\rangle=4D$. The approximation by white noise in Eq. \eqref{eq:potential} does not preserve correlations around multiples of $\tau$, like those shown in Fig. \ref{fig:rtd}(a). The potential $V(x)$ is shown in Fig. \ref{fig:kura1pot}. It is a parabolic potential modulated by a cosine function. The local minima $x_k=\omega_k\tau$ correspond to the frequencies in the noise-free case $D=0$. Our reduction procedure does not affect them and their calculation by the potential extrema reveals Eq.~\eqref{eq:omega}. The local maxima $x_m$ correspond to unstable solutions of the deterministic system.

The phase difference $x(t)$ relates in a simple way to the frequency measure $x(t)/\tau=\omega(t)$. Hence, the stationary distribution of frequencies $p(\omega)$ is given by a Boltzmann factor \cite{Kampen}
\begin{equation}
p(\omega)\propto e^{-\frac{V(\omega\tau)}{2D}}=e^{-\frac{\tau}{4D}(\omega-\omega_0)^2}e^{\frac{\kappa\cos\omega\tau}{2D}}\label{eq:frequencydistribution}\,.
\end{equation} 

We recognize a Gaussian envelope with mean $\omega_0$ and variance $\sigma^2=2D/\tau$. This envelope corresponds to the probability distribution of a random walk. Thus, while the total frequency range is given by $2\kappa$, the range of visited frequencies scales with $\sqrt{D/\tau}$. As the spacing between the orbits scales inversely with the feedback delay, the number of attended orbits grows as $\sqrt{D\tau}$.  

The coupling function, which appears in the second factor of Eq. \eqref{eq:frequencydistribution}, determines the location and the shape of the different peaks. As the feedback strength increases, the peaks in the distribution become more pronounced. In Fig. \ref{fig:var} we compare our analytical result for the simplified system (Eq. \eqref{eq:frequencydistribution}) with numerical simulations of the original delay system (Eq. \eqref{eq:model}). We find that our theoretical results provide an excellent approximation for the distribution of frequencies and thus prove the validity of the applied reduction method.

Also the average residence times can be approximated by the potential model \eqref{eq:potential}: In the limit of low noise, the escape rates from an orbit with frequency $\omega_k$ are given by the Kramers rate \cite{Kra40,McN89}

$$r_{\pm}(\omega_k)= \frac{\sqrt{-V''(\omega_k\tau)V''(x_m)}}{2\pi}e^{-\frac{\Delta V}{2D}}\,,$$
\noindent where the suffix denotes whether the oscillator hops to a mode with a higher or a lower frequency. The average residence time $T_0(\omega_k)$ reads then
$$T_0(\omega_k)\approx\frac{1}{r_+(\omega_k)+r_-(\omega_k)}\,.$$
For strong coupling and large feedback delay $\kappa\tau\gg 1$, a multitude of orbits are stable, with $\omega_k\tau\approx 2n\pi$ and $x_m=(2n+1)\pi$. The average residence time is then further approximated as
\begin{equation}
T_0(\omega_k)\approx\frac{\pi}{\kappa}\frac{e^{\frac{\kappa}{D}+\frac{\pi^2}{4\tau D}}}{\cosh\left(\frac{\pi(\omega_k-\omega_0)}{2D}\right)}\label{eq:rtd1}\,.
\end{equation}
\noindent We compared the average residence time of the different periodic orbits with our theoretical result (Eq. \eqref{eq:rtd1}) in Fig. \ref{fig:rtd}, and the approximation gives good results. Consequently, the average residence time $T_0(\omega_k)$ increases with the feedback strength $\kappa$, which determines the depth of the potential wells, and decreases with the noise strength $D$. For a fixed frequency $\omega_k$ the feedback delay $\tau$ has a limited influence on the residence times, for long delays the delay dependency even vanishes. 

Only the orbits with a frequency close to the natural frequency have a considerable average residence time, and are in this sense robust to noise. This range of these frequencies scales with $D$, and does not depend on the delay time or the coupling strength. Due to the frequency difference of $2\pi/\tau$, the number of orbits that is robust to noise scales approximately as $D\tau$. Moreover, there is a difference in the mode hopping behavior at long and short delay times: For long delays, the $\sqrt{D/\tau}$-range of attended orbits is much smaller than the range of robust frequencies, so that all visited orbits have a similar average residence time. If the delay is shorter, as it is the case for our choice of parameters, significant differences in the residence times of the orbits are observed.


\section{Two mutually coupled phase oscillators}

\begin{figure}
\includegraphics[width=0.49\columnwidth]{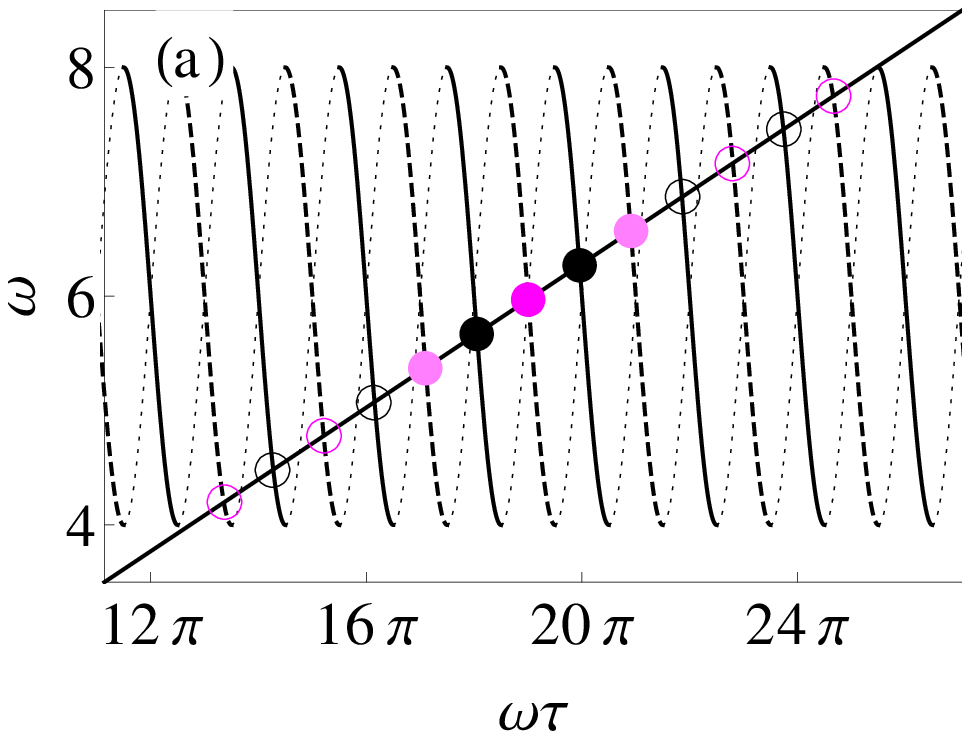}
\includegraphics[width=0.49\columnwidth]{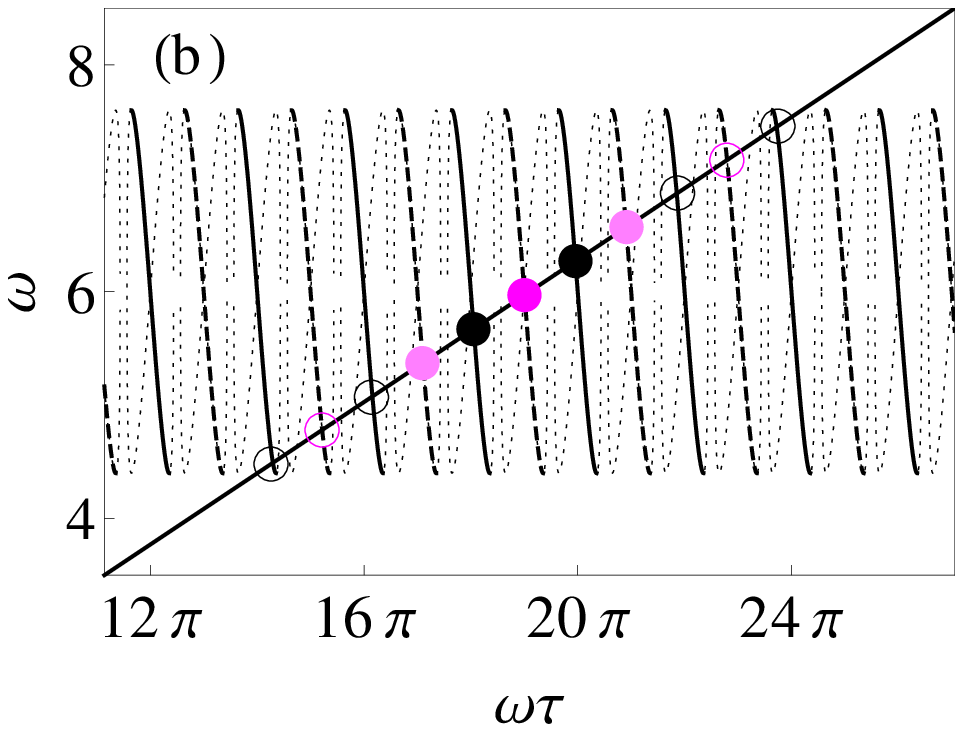}
\caption{(Color online) Graphical determination of the locking frequencies of two (a) identical and (b) nonidentical delay-coupled phase oscillators. Intersection with the thick full (dashed) line correspond stable in-phase (anti-phase) orbits. The filling of the black (magenta) circles relates to the relative probability that the in-phase (anti-phase) orbit is visited by the corresponding stochastic system, darker labeling corresponds to a higher probability to find a frequency $\omega(t)\approx \omega_k$. The corresponding probability distributions are shown in Fig. \ref{fig:var} (b,c). In panel (b) a stable orbit is labeled as in-phase if $-\pi/2<\delta<\pi/2$. Just like for a single feedback oscillator, the frequencies $\omega_k\approx\omega_0$ are most often visited, the width of the frequency distribution is however smaller. Parameters are $\omega_0=6$, $\kappa=2$, $\tau=10$ and (b) $\Delta=0.8$.}
\label{fig:kura2}
\end{figure}

More common than a single oscillator driven by its own delayed feedback are coupled oscillators. We consider here the simple case of two mutually delay-coupled oscillators with independent noise sources. This system is modelled by
\begin{eqnarray}
\dot{\phi}_1(t) & = & \omega_{01}+\kappa\sin(\phi_{2}(t-\tau)-\phi_1(t))+\xi_1(t)\nonumber\\
\dot{\phi}_2(t) & = & \omega_{02}+\kappa\sin(\phi_{1}(t-\tau)-\phi_2(t))+\xi_2(t)\label{eq:Zn}\,,
\end{eqnarray}
with $\omega_{01,02}=\omega_0\mp\Delta/2$, and $\Delta$ being the detuning between the oscillators. We repeat first the case of identical oscillators ($\omega_{01}=\omega_{02}\equiv \omega_0$) without noise ($D=0$) \cite{AmpPhase,ikke}. The system does not only have in-phase synchronized oscillations $\phi_1(t)=\phi_2(t)=\omega_k t$, but also anti-phase synchronized orbits $\phi_1(t)=\phi_2(t)+\pi=\omega_k t$. The frequencies of the in-phase orbits are identical to the single feedback system; they are given by $\omega_k=\omega_0-\kappa\sin(\omega_k\tau)$. For the anti-phase orbits the frequencies can be found by solving $\omega_k=\omega_0+\kappa\sin(\omega_k\tau)$. The coupled system thus has twice as many coexisting periodic orbits as the single system. In-phase orbits are stable for $\cos(\omega_k\tau)>0$ and anti-phase orbits for $\cos(\omega_k\tau)<0$. A graphical determination of the frequencies is shown in Fig. \ref{fig:kura2}(a): stable in-phase and anti-phase frequencies alternate each other. For large $\kappa\tau\gg1$ the frequencies $\omega_k$ close to the natural frequency $\omega_0$ are approximated as $\omega_k\tau\approx n\pi$, so that the separation between the frequencies approaches $\pi/\tau$.

Without noise, nonidentical oscillators still synchronize to a common frequency if the coupling is strong enough $|\Delta|<2\kappa$ \cite{sch89.1}. Detuned oscillators, however, are no longer exactly in-phase or anti-phase with each other, but they exhibit a phase difference $\delta$ depending on the locking frequency and the detuning. We find for the frequencies $\omega_k$ and the phase difference $\delta$
\begin{eqnarray}
\omega_k &= & \omega_0 - \kappa\sin(\omega_k\tau)\cos\delta\nonumber\\
\sin\delta & = & \frac{\Delta}{2\kappa\cos(\omega_k\tau)}\label{eq:omegadet}\,,
\end{eqnarray}
if the conditions $\cos(\omega_k\tau+\delta)>0$ and $\cos(\omega_k\tau-\delta)>0$ hold, an orbit is stable. We solve Eq. \eqref{eq:omegadet} graphically in Fig. \ref{fig:kura2}(b). For nonidentical oscillators the frequency range is reduced to $2\kappa-\Delta$; for large $\kappa\tau$ however neither the locking frequencies $\omega_k$ nor their stability is largely affected by the detuning.

\begin{figure}
\includegraphics[width=\columnwidth]{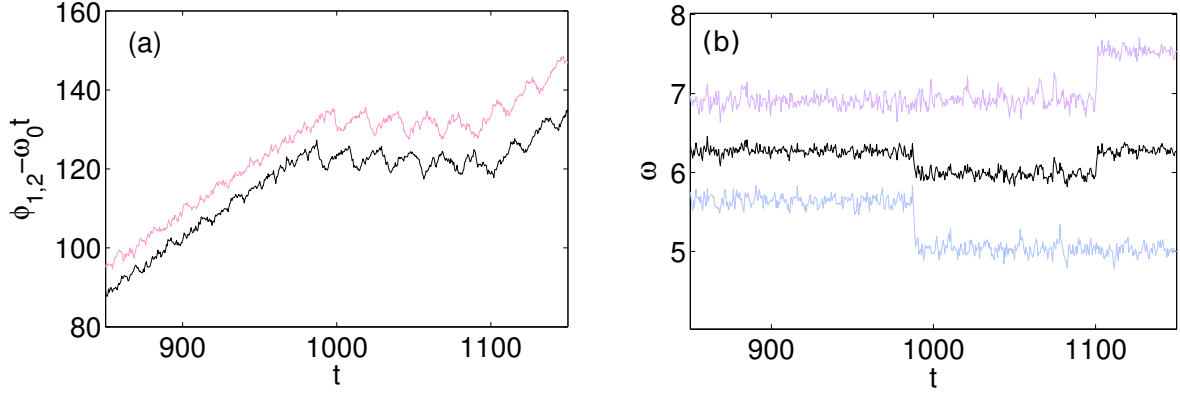}
\caption{(Color online)(a) The phase evolutions $\phi_{1}-\omega_0t$ (pink) and $\phi_{2}-\omega_0t$ (black) of two identical noisy Kuramoto oscillators coupled with delay. We subtracted the natural frequency $\omega_0$ for better visibility of the mode hoppings.
(b) The time evolution of the frequency $\omega(t)=(\phi_1(t)+\phi_2(t)-\phi_1(t-\tau)-\phi_2(t-\tau))/(2\tau)$ for two coupled oscillators (black), together with the phase differences $x_1(t)/\tau=(\phi_1(t)-\phi_2(t-\tau))/\tau$ (upper pink curve) and $x_2(t)/\tau=\phi_2(t)-\phi_1(t-\tau)$ (lower blue curve). The dashed lines indicate the mode hoppings. Parameters are $\omega_0=6$, $\kappa=3$, $\tau=10$ and $D=0.5$.}
\label{fig:kura2tt}
\end{figure}

We show the phase evolution of two identical delay-coupled oscillators in Fig. \ref{fig:kura2tt}. Mode hopping happens in two stages: if one oscillator, the leader, changes its frequency, the other oscillator, the laggard, follows a delay time later. Looking at the evolution of the driving terms $\phi_{1,2}(t)-\phi_{2,1}(t-\tau)$, shown in Fig. \ref{fig:kura2tt}(b), it is clear that during a transition the driving term of the leader changes with $2\pi$, while the laggard changes its frequency without a phase jump in its drive. As a frequency measure for the coupled system we use the mean frequency of the two oscillators averaged over the past delay interval $\omega(t)=(\phi_1(t)+\phi_2(t)-\phi_1(t-\tau)-\phi_2(t-\tau))/(2\tau)$; we thus capture the frequency transition of the leading oscillator. The two oscillators initiate equally many transitions, hence the role of leader and laggard changes randomly. For nonidentical oscillators, however, if the system speeds up, the fast oscillator is more often the leader, while if the oscillators slow down, the slow oscillator is leading the dynamics.

\begin{figure}[t]
\includegraphics[width=\columnwidth]{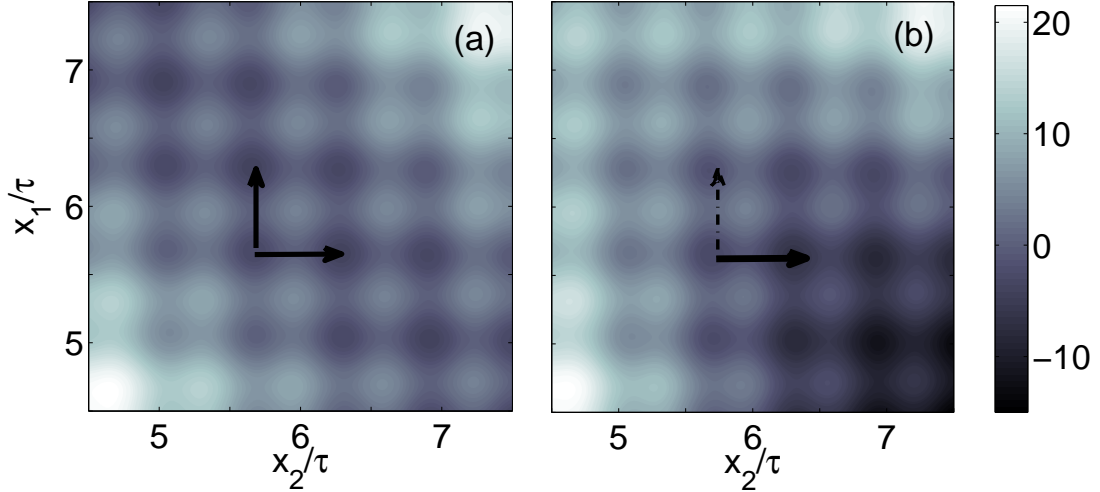}
\caption{Twodimensional potential for two coupled Kuramoto oscillators, without (a) and with (b) detuning. The arrows indicate the two pathways for a transition between two frequencies, thicker arrows correspond to more probable pathways. Parameters are $\omega_{0}=6$, $\kappa=2$, $\tau=10$ and (b) $\Delta=0.8$}
\label{fig:potential-det}
\end{figure}

Also for mutually coupled oscillators it is possible to define a delay-free Langevin formalism. We rewrite the system as a function of the driving terms $x_1(t)$ and $x_2(t)$, defined as $x_{1,2}(t)=(\phi_{1,2}(t)-\phi_{2,1}(t-\tau))$. We then assume that the oscillators are locked to the same fixed frequency over the delay interval, and as such, that $\dot{\phi}_1(t-\tau)$ and $\dot{\phi}_2(t-\tau)$ only differ in the contribution of the noise. This leads to the main reduction
\begin{equation}
\dot{\phi}_{1,2}(t-\tau)\approx(x_1(t)+x_2(t))/(2\tau)+\xi_{1,2}(t-\tau)\,.
\end{equation}
In this way we can rewrite the system as a function of a twodimensional potential: 

\begin{eqnarray}
\dot{x}_{1,2}(t)&=&-\frac{\partial V_{1,2}}{\partial x_{1,2}}+\tilde{\xi}_{1,2}(t) \mbox{ with }\nonumber\\
V(x_1,x_2) &=& \frac{1}{4\tau}(2x_0-x_1-x_2)^2+\frac{\Delta}{2}(x_1-x_2)\nonumber\\
					 & & -\kappa\left(\cos x_{1}+\cos x_2\right) \label{eq:potential-det}\,,
\end{eqnarray}%
with $x_0=\omega_0\tau$ and $\tilde{\xi}_{1,2}(t)=\xi_{1,2}(t)-\xi_{2,1}(t-\tau)$. This potential is shown in Fig. \ref{fig:potential-det}. The wells are located at $(x_1,x_2)=(\omega_k\tau +2n\pi-\delta,\omega_k\tau-2n\pi+\delta)$. The frequency of the system is then given by the average frequency $\omega=(x_1+x_2)/(2\tau)$. As the phase difference between the oscillators is only determined upon a multiple of $2\pi$, the potential is $4\pi$-periodic with respect to $x_1-x_2=x_A$. For identical oscillators there are thus two equally probable pathways for a transition: $x_1$ changes with almost $2\pi$, while $x_2$ remains almost constant, and $\phi_1(t)$ leads the dynamics, and vice versa. These pathways are indicated by arrows in Fig. \ref{fig:potential-det}. Transitions typically take place between orbits with a minimal frequency difference, and therefore with a different oscillation pattern. If the oscillators are identical, we obtain the frequency distribution $p(\omega)$ by integrating over the phase difference $x_A$. We find
\begin{eqnarray}
p(\omega) & \propto & \int_0^{4\pi}dx_A e^{-\frac{V(\omega\tau,x_A)}{2D}}\nonumber\\
 			& \propto & e^{-\frac{\tau}{2D}(\omega-\omega_0)^2}I_0(\kappa\cos\omega\tau/D)\label{eq:freqdistr2}\,,
\end{eqnarray} 
with $I_0(y)$ being the modified Bessel function of the first kind, $I_0(y)=\sum \frac{y^{2n}}{2^{2n}(n!)^2}$.

Like for the single oscillator, the frequency distribution can be written as a Gaussian envelope multiplied with a factor determining the separate peaks. However, the variance of the envelope decreases with a factor $1/2$ compared to the single feedback system. The Bessel function $I_0(y)$ is symmetric: we find alternating peaks corresponding to in-phase and anti-phase orbits, their height only depends on their respective frequencies $\omega_k$ and not on the oscillation pattern. We compare our numerical and theoretical results for the frequency distribution in Fig. \ref{fig:var}(b). The agreement is excellent. The oscillators are thus always synchronized (except during the delay interval following a transition), but they spend a proportion of time in in-phase as well as in anti-phase orbits. As a result, for long enough delays, the overall correlation between the oscillators vanishes at zero lag, but shows maxima at odd multiples of the coupling delay.

For nonzero detuning $\Delta>0$, the potential (Eq. \eqref{eq:potential-det}) is tilted, as is shown in Fig. \ref{fig:potential-det}(b). Consequently the phase difference $x_A(t)$ between the oscillators preferentially increases during a mode hopping. The most probable, and the least probable transition pathway between two frequencies are also sketched on Fig. \ref{fig:potential-det}(b). The ratio between the transition rates is approximated by $exp(-pi*Delta/2D)$, so that for large detunings it is reasonable to assume that all the transitions to a higher frequency are induced by the faster oscillator $x_2$, and the transitions to a lower frequency by the slower one $x_1$. For $\kappa\tau$ sufficiently large, we can approximate the envelope by assuming detailed balance
\begin{eqnarray}
p(\omega_k)r_+(\omega_k) & = & p(\omega_{k+1})r_-(\omega_{k+1})\Leftrightarrow\nonumber\\
\frac{p(\omega_{k+1})}{p(\omega_k)} & \approx & e^{-\frac{\Delta V(\omega_{k+1}\rightarrow \omega_k)-\Delta V(\omega_{k}\rightarrow \omega_{k+1})}{2D}}\nonumber\\
& \approx & e^{-\frac{\tau}{2D}\left((\omega_{k+1}-\omega_0)^2-(\omega_{k}-\omega_0)^2+\frac{2\delta}{\tau}(2\omega_0-\omega_{k+1}-\omega_k)\right)}\nonumber\\
                                    & \approx & e^{-\frac{\tau}{2D}\left(1-\frac{2\delta}{\pi}\right)\left((\omega_{k+1}-\omega_0)^2-(\omega_k-\omega_0)^2\right)}\,.
\end{eqnarray}
This corresponds to a Gaussian envelope of the frequency distribution with mean $\omega_0=(\omega_1+\omega_2)/2$ and variance $\sigma^2=D/(\tau(1-\epsilon))$, with $\epsilon=2\arcsin(\Delta/2\kappa)/\pi>0$. The distribution of frequencies thus becomes broader due to the detuning, in agreement with the numerical results for the full delay system. In Fig. \ref{fig:var}(c) we show the approximated Gaussian envelope together with the simulated distribution of frequencies.

For identical oscillators, we approximate the residence times of the orbits by assuming all the transitions take place via the two optimal pathways. We obtain then for the mean residence time
\begin{equation}
T_0(\omega_k)=\frac{1}{2r_+(k)+2r_-(k)}\approx\frac{\pi}{2\kappa}\frac{e^{\frac{\kappa}{D}+\frac{\pi^2}{8\tau D}}}{\cosh\left(\frac{\pi(\omega_k-\omega_0)}{2D}\right)}\label{eq:rtd2}\,.
\end{equation}
This corresponds to half of the lifetime of the orbits of a single oscillator with the a roundtrip delay $2\tau$. We compare numerical and theoretical results in Fig. \ref{fig:rtd}(b).


\section{Extension to a ring of Kuramoto oscillators}

It is possible to extend these results to a unidirectional ring of $N$ oscillators. Such system is then modelled by
\begin{equation}
\dot{\phi}_n(t)=\omega_0 + \Delta_n + \kappa\sin(\phi_{n+1}(t-\tau)-\phi_n(t))+\xi_n(t)\label{eq:uniring}\,,
\end{equation}
with $N+1\equiv 1$. Without detuning, the coupling topology allows for in-phase oscillations $\phi_n(t)=\omega_k t$ and several out-of-phase oscillation patterns $\phi_n(t)=\omega_k t+n\Delta\phi$, with $\Delta\phi=2m\pi/N$. The corresponding frequencies are given by $\omega_k=\omega_0-\kappa\sin(\omega_k\tau-\Delta\phi)$, and they are stable if $\cos(\omega_k\tau-\Delta\phi)>0$ \cite{ikke}. This results in alternating orbits with a different oscillation pattern. For strong coupling and long delay, the frequencies are separated by $2\pi/(N\tau)$.

Defining $x_n=\phi_n(t)-\phi_{n+1}(t-\tau)$, and assuming that the instantaneous frequencies of the oscillators can be approximated by the mean frequency averaged over the delay interval and their noise source,
$$\dot{\phi}_n(t-\tau)\approx\frac{1}{N\tau}\displaystyle\sum\limits_{l=1}^N x_l +\xi_n(t-\tau)\,,$$
we find an N-dimensional potential
\begin{equation}
V(x_1,\hdots, x_N)=\frac{N}{2\tau}(x_0-x_S)^2 +\displaystyle\sum\limits_{l=1}^N\Delta_n x_n + \kappa\displaystyle\sum\limits_{l=1}^N\cos x_l\label{eq:potentialuni}\,,
\end{equation}
with $x_S=\frac{1}{N}\displaystyle\sum\limits_{l=1}^N x_l$. The frequency of the system is then measured by $\omega(t)=x_S(t)/\tau$. It is no longer possible to compute the frequency distribution $p(\omega)$ in terms of simple analytical expressions as above. However, for identical oscillators ($Delta_n=0$) it is straighforward to see that the parabolic term in Eq. \eqref{eq:potentialuni} leads to a Gaussian envelope. The variance of this envelope is given by $\sigma^2=2D/(N\tau)$, and thus scales inversely with the total roundtrip delay $N\tau$. As the frequency difference between the orbits is approximated by $2\pi/(N\tau)$, the number of attended orbits scales as $\sqrt{N\tau}$.  Moreover, the potential is symmetric with respect to the different oscillation patterns, so that each pattern is equally often visited in the long delay limit. 

For low noise, zero detuning and large $\kappa\tau$, we find that the average residence times scale inversely with the number of oscillators in the ring, and depend weakly on the total roundtrip delay. They are approximated by
\begin{equation}
T_0(\omega_k)=\frac{1}{Nr_+(\omega_k)+Nr_-(\omega_k)}\approx\frac{\pi}{N\kappa}\frac{e^{\frac{\kappa}{D}+\frac{\pi^2}{4N\tau D}}}{\cosh\left(\frac{\pi(\omega_k-\omega_0)}{2D}\right)}\label{eq:rtdN}\,.
\end{equation}

We compared the frequency distributions and residences times of the simplified non-delay system with simulations of three, four and five delay-coupled oscillators, and the agreement is excellent (not shown).


\section{General periodic systems with delayed coupling}

In order to investigate whether our results are valid in a broader context, we compare the switching behavior of other nonlinear delay-coupled oscillators to our results for phase oscillators. The Kuramoto model is a weak-coupling limit, which only describes the phase dynamics, and does not take any influence on the amplitude into account; therefore we expect that our results mainly apply for weak coupling.

First, we sketch the deterministic periodic solutions in a general delay system. For a single oscillator, it is known that a feedback delay induces coexisting periodic orbits, with a frequency separation of $2\pi/\tau$ \cite{Yanchuk09}. We show here briefly that in a unidirectional ring of identical oscillators, a delay gives rise to alternating in-phase and out-of-phase orbits, in a similar way as for phase oscillators. For general limit cycle systems, unlike for phase oscillators, it is not so straightforward to determine the respective orbits and their stability properties.

Extending the approach of Yanchuk and Perlikowski for a single feedback system \cite{Yanchuk09}, we consider a set of $N$ identical nonlinear systems coupled in a unidirectional ring with delay 
\begin{equation}
\dot{x}_n(t)=f(x(t),x_{n+1}(t-\tau))\label{unialg}\,.
\end{equation}
where $x_{N+1}\equiv x_1$. For the following we assume, that this network allows for an in-phase synchronized periodic solution $x_n(t)=x_{n-1}(t)=x_n(t+T)$ at a coupling delay $\tau=\tau_0$. Shifting the delay to $\tau_1=\tau_0+T/N$, the same periodic orbit is a solution of the system, the oscillators however exhibit a phase difference $x_n(t)=x_{n+1}(t-T/N)=x_n(t+T)$. Similarly, we find the same waveform appearing with all the other out-of-phase patterns that are allowed in the ring: a pattern corresponding to $x_n(t)=x_{n-1}(t-kT/N)=x_n(t+T)$ can be found at a delay $\tau_k=\tau_0+kT/N$. An orbit with a period $T$ thus reappears when shifting the delay by an amount $T/N$.

The periodic solutions are organized in branches: as the delay increases, the period $T$ of an orbit varies continuously between a minimal period $T_{\min}$ and a maximal period $T_{\max}$. For a fixed delay $\tau$, the number of coexistent orbits resulting from a single branch can then be estimated in the following way: we have $\tau\approx nT_{\max}/N\approx mT_{\min}/N$. The number of periodic states is then estimated as $m-n=N\tau(T_{\min}^{-1}-T_{\max}^{-1})$, with in-phase and out-of-phase orbits alternating each other. The frequency difference between two orbits is approximated by $2\pi/(N\tau)$, just like for phase oscillators. It is possible to show that the stability of these orbits depends on their period, but not on the oscillation pattern. In the long delay limit the stability no longer depends on the number of oscillators in the ring, or the coupling delay.

\begin{figure}[t]
\includegraphics[width=0.49\columnwidth]{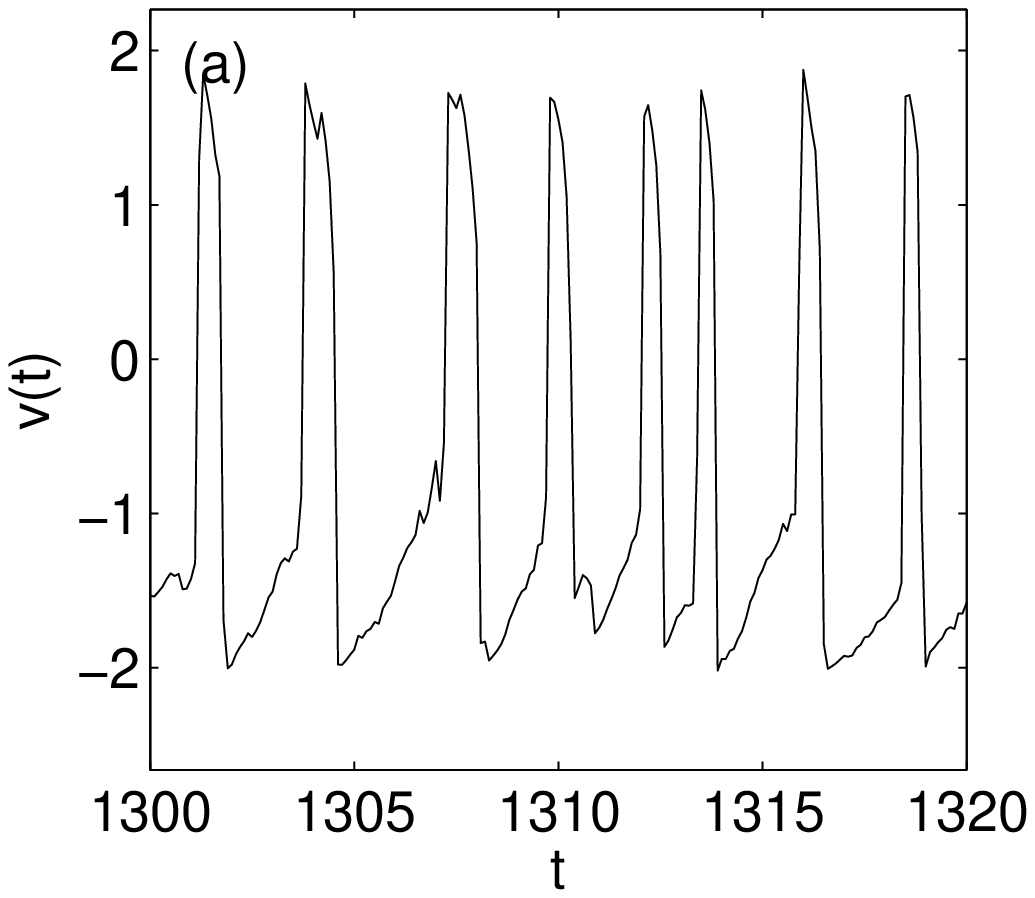}
\includegraphics[width=0.49\columnwidth]{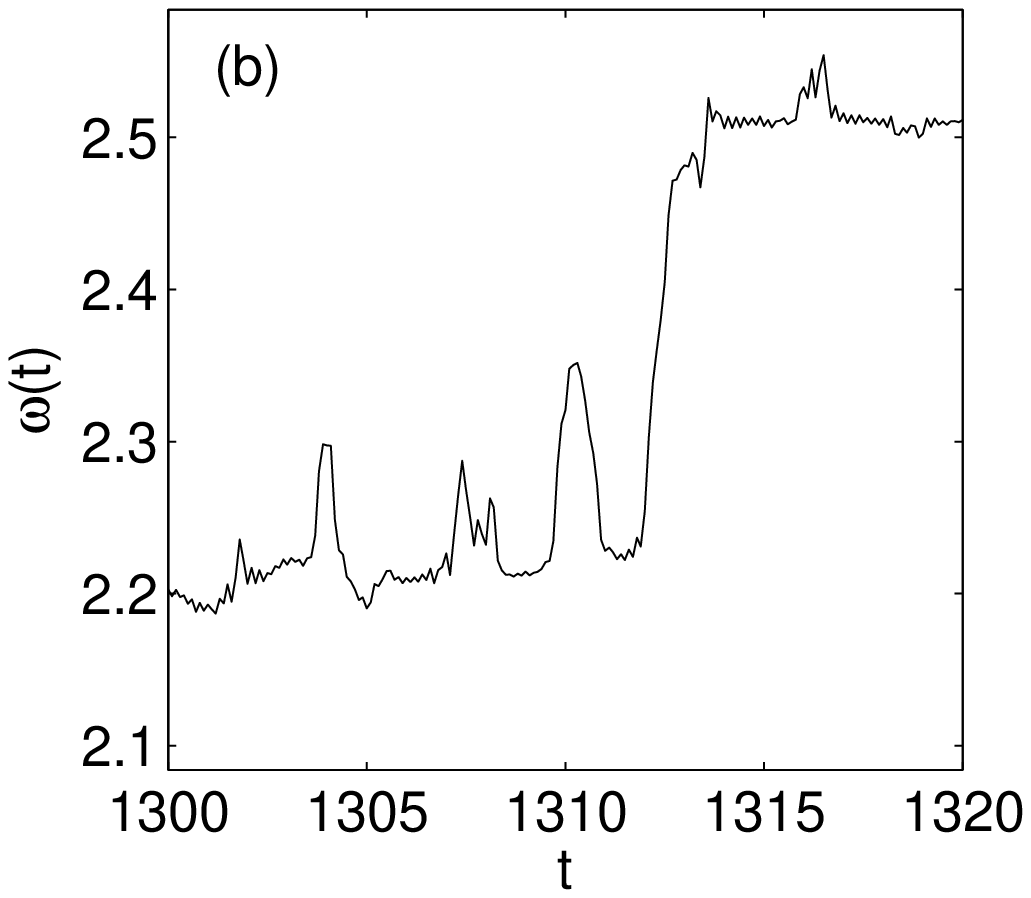}
\caption{(a) Time trace and (b) frequency evolution of a FitzHugh-Nagumo oscillator with feedback. Parameters are $\epsilon=0.01$, $a=0.9$, $k=0.2$, $\tau=20$ and $\tilde{D}=0.0145$.}
\label{fig:FHNtt}
\end{figure}

As an examplary system, we investigate numerically stochastic switching between such coexistent orbits in FitzHugh-Nagumo oscillators. We simulated a single oscillator ($N=1$) with delayed feedback, and two identical mutually delay-coupled oscillators ($N=2$). Our oscillator is modelled by
\begin{eqnarray}
\epsilon\dot{v}_n(t) & = & v(t)-\frac{{v_n}^3(t)}{3}-w_n(t)+k(v_{n+1}(t-\tau)-v_n(t))\nonumber\\
\dot{w}_n(t) & = & v_n(t)+a +\xi_n(t)\,,
\end{eqnarray}
with $(v_{N+1},u_{N+1})\equiv (v_1,u_1)$, and $\xi_n(t)$ being Gaussian white noise with a variance given by $\langle\xi^2(t)\rangle=2\tilde{D}$. We choose our parameters so that without delayed coupling and without noise the oscillator(s) show periodic spiking dynamics. A typical timetrace of an oscillator with noise and feedback, which performs a mode hopping, is shown in Fig. \ref{fig:FHNtt}(a).

We analyze the mode hopping in a similar way as for phase oscillators. We define the phase of the oscillators by the Hilbert-transform of the fast variable, $\phi_n(t)=\arg(\mathcal{H}(v_n(t))$, but similar results were obtained by using the alternative definition $\phi_n(t)=\arctan(v_n(t)/w_n(t))$. In both cases the waveform is very different from sinusoidal, so the frequency shows large fluctuations within a period even without noise. The frequency measure is given by $\omega(t)=\sum(\phi_n(t)-\phi_n(t-\tau))/(N\tau)$, similar as for Kuramoto oscillators. We show the evolution of the frequency $\omega(t)$ in Fig. \ref{fig:FHNtt}(b). Although the mode hopping event is hard to detect in the timetrace (Fig. \ref{fig:FHNtt}(a)), it is clearly visible in the frequency. Comparing Figs. \ref{fig:FHNtt}(b) and \ref{fig:kura1tt}(b), we find that the irregular waveform of the FitzHugh-Nagumo oscillator results in larger and asymmetric excursions from the deterministic frequency.

\begin{figure}[t]
\includegraphics[width=\columnwidth]{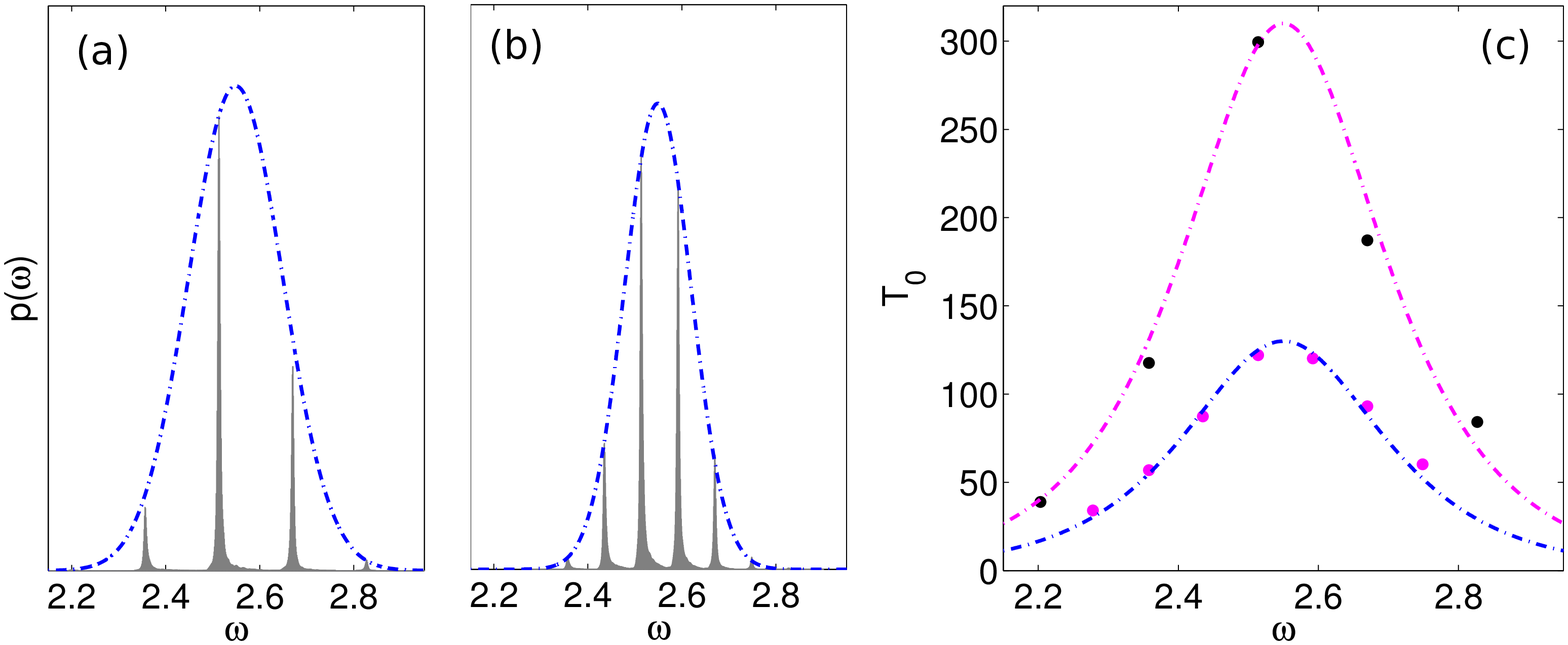}
\caption{(Color online) Frequency distribution $p(\omega)$ for one (a) and two (b) coupled FitzHugh-Nagumo oscillators with delay, with $\epsilon=0.01$, $a=0.9$, $k=0.2$, $\tau=40$ and $\tilde{D}=0.0145$. The blue dashed curve shows the Gaussian envelope for the corresponding Kuramoto oscillator(s) with $\omega_0=2.55$, $D=0.2$ and $\tau=40$. In panel (c) the corresponding average residence times are shown for one (upper black dots) and two (lower pink dots) oscillators, the dashed curves represent the Kuramoto approximation for one (Eq. \eqref{eq:rtd1}, upper pink curve) and two (\eqref{eq:rtd2}, lower blue curve) elements for $\omega_0=2.55$, $D=0.2$, $\tau=40$ and $\kappa=0.815$}
\label{fig:FHN}
\end{figure}

Fig. \ref{fig:FHN} (a,b) compares the frequency distributions $p(\omega)$ of a single oscillator with feedback, and two mutually coupled oscillators. For the single oscillator, shown in Fig. \ref{fig:FHN}(a), we find five different peaks, separated by a frequency difference of $2\pi/\tau$. The shape of the different peaks is asymmetric; this feature results from the asymmetric waveform of the spikes. The frequency distribution $p(\omega)$ for two mutually coupled oscillators is shown in Fig. \ref{fig:FHN}(b): we find the peaks at the same frequency as for the single element; they correspond to in-phase orbits. Between the in-phase peaks we find maxima that can be associated to anti-phase orbits. Just like for phase oscillators, the frequency distribution for the coupled system has the same mean, and half of the variance as the single system. The corresponding average residence times of each orbit are shown in Fig. \ref{fig:FHN}(c). The average residence times of the single oscillator (black dots) are larger than those of the coupled system (pink dots). Moreover, the average residence times show the same trend as for coupled phase oscillators: the orbits with a central frequency are most robust against noise.

The agreement with the Kuramoto oscillators is even quantitative: In Fig. \ref{fig:FHN}(a,b) we compared the frequency distributions with Gaussians (blue dashed lines), which have the same mean and variance as the original distributions; the maxima of the peaks approximately lie on this Gaussian curve, both for the single feedback oscillator and the two delay-coupled oscillators. From the mean and variance, we identify the natural frequency $\omega_0$ and the noise strength $D$ of the corresponding Kuramoto model with the same delay $\tau$. The coupling phase can be found by the position of the in-phase peaks $\theta\approx \omega_k\tau$. 

We also compared in Fig. \ref{fig:FHN}(c) the Kuramoto residence times for a single oscillator (Eq. \eqref{eq:rtd1}, pink dashed curve) and for two coupled oscillators (Eq. \eqref{eq:rtd2}, blue dashed curve) to the residence times for FitzHugh-Nagumo elements. We thereby used the parameters $D$ and $\omega_0$ determined from the frequency distributions, the coupling strength $\kappa$ can then be estimated from the average residence times. We find that, for the same coupling strength $\kappa$ for the single and the coupled system, the residence times are well approximated by the phase model. Moreover, we find a single parameter set $(\omega_0,\kappa,D,\theta)$ which models the frequency distribution and the average residence times for both a single feedback and two coupled oscillators. Hence, the scaling properties of the stochastic periodic dynamics with the delay time and the oscillator number are reproduced. 

\section{Summary and discussion}

We have studied the influence of additive noise on one, two and a ring of phase oscillators coupled with delay. In such systems multiple periodic orbits coexist, and under influence of noise the oscillators hop from one orbit to another. We approximated the system as a noisy particle in potential well; both for the distribution of frequencies as for residence times the approximation is excellent. Although our approximation only applies for weak noise, we obtain a good agreement for the distribution of frequencies for all noise strenghts.

However, it should be remarked that in the simplified model some dynamical phenomena are not reproduced. The most prominent example are the delay stochastic resonance peaks in the residence time distribution, shown in Fig. \ref{fig:rtd}(a). Also frequency oscillations with a periodicity of a roundtrip delay time, which are typically present in the system, are no longer visible. Transient behavior is different as well: in the delay system a transient decays at a rate proportional to the inverse delay time, while in the reduced system the decay happens much faster.

We found that the oscillators only visit a fraction of the deterministic stable orbits: whereas the number of deterministic orbits scales with oscillator number $N$, delay time $\tau$ and coupling strength $\kappa$, the number of visited orbits scales as $\sqrt{DN\tau}$, and does not depend on the coupling strength $\kappa$. The orbit with a frequency closest to the natural frequency is the most probable, irrespective of its oscillation pattern.  

Our results on the average residence times indicate the robustness of the orbits against weak noise. The most robust orbits are those with a frequency close to the natural frequency, also irrespective of the oscillation pattern. The sensitivity of an orbit against noise depends strongly on the coupling strength, the coupling delay plays only a minor role. The number of robust orbits scales as $DN\tau$.

For two delay-coupled oscillators, and for unidirectional rings the systems does not show any preference for a particular oscillation pattern. The different oscillation patterns are equally often attended, in the long delay limit. However, this symmetry between in-phase and out-phase patterns depends on the coupling topology. We also simulated three, four and five delay-coupled Kuramoto oscillators in an all-to-all configuration. In this case however a description as a noisy particle in a potential is not accurate, as not only periodic dynamics is observed. The distribution of frequencies looks different: the peaks associated to in-phase orbits are considerably higher than those corresponding to out-of-phase dynamics. For long delays even only in-phase periodic orbits are visible in the frequency distributions. We find that the frequency distributions narrows with the number of oscillators: the variance of the frequency distribution scales as $1/(N-1)$.

The Kuramoto model is a weak coupling approximation for limit cycle oscillators. Therefore we expect our results to apply for delay-coupled nonlinear systems showing stable periodic dynamics. In particular, the Kuramoto approximation applies when the coupling mainly influences the oscillation phase, while the waveform or oscillation amplitude is hardly affected. We found indeed a good correspondence between Kuramoto and FitzHugh-Nagumo oscillators in this case. However, we expect the approximation to break down as the coupling strength increases and amplitude instabilities play a role in the dynamics.

Not only in stable oscillatory systems, but also in a chaotic attractor a delay has the effect of inducing multiple periodic orbits. Hence, the chaotic attractor of two delay-coupled chaotic systems contains in-phase as well as anti-phase orbits, and they have similar stability properties (for long enough delay). Therefore, it is not surprising that we find the same correlation pattern, with a high correlation at the delay time, but no correlation at zero lag, for coupled noisy oscillators and chaotic systems with delay \cite{hei01.1}. However, chaotic and stochastic systems show different scaling behavior with the delay time and the number of coupled elements.

It is worth noting that the envelopes of the frequency distributions are the same as those for a random walk. The delayed feedback only imposes restrictions on the distribution of the two point distribution of $x(t)=\phi(t)-\phi(t-\tau)$, but it does not affect the envelope. On timescales much shorter than the delay $t_0\ll\tau$, the influence of the feedback is even not visible: the two point distribution of $\phi(t)-\phi(t-t_0)$ is identical to the one of a random walk. A possible explanation of this surprising phenomenon lies in the fact, that the equations of motion do not impose any restrictions on this phase difference, as long as $t_0$ is different enough from $\tau$. Hence, the random-walk can explore the possible range. However, on timescales equal or larger than the delay, the dynamics (i.e. the timetrace) of an oscillator with delayed feedback differs significantly from a random walk. Also the two point distributions show a clear fingerprint of the delay time, and for $t_0>\tau$, a larger variance than a random walk. We believe that the issue of two-/$N$-point distributions in delay systems is worth being studied in more detail.


\begin{acknowledgements}
This work was initiated during a visit at TU Berlin, and O.D. is grateful to Eckehard Sch\"oll and Andrea V\"ullings for their hospitality and many ideas. Furthermore, O.D. thanks Ido Kanter and Jordi Zamora for fruitful discussions. T.J. acknowledges support by a fellowship within the Postdoc-Programme of the German Academic Exchange Service (DAAD).
\end{acknowledgements}

\end{document}